\documentclass[reprint,aps,prc,twocolumn,superscriptaddress,floatfix,amsmath,amssymb,10pt,longbibliography]{revtex4-2}

 \usepackage{graphicx}
\usepackage{bm}
\usepackage{amsmath} \usepackage{braket} \usepackage{epsfig}
\usepackage{tensor}
\usepackage{multirow}
\usepackage[T1]{fontenc}
\usepackage[version=4]{mhchem}
\usepackage{titlesec}
\usepackage{ifthen}
\usepackage{sidecap}
\usepackage{listings}
\usepackage[para,online,flushleft]{threeparttablex}
\usepackage{algorithm}
\usepackage{algpseudocode}
\usepackage{comment}
\usepackage{MnSymbol}
\usepackage{xcolor}
\usepackage{xr-hyper}
\usepackage{hyperref}
\hypersetup{breaklinks=true,colorlinks=true,linkcolor=blue,citecolor=blue,filecolor=magenta,urlcolor=cyan}

\usepackage{orcidlink}

\usepackage[all]{hypcap}
\newcommand{\subref}[2]{\hyperref[#1]{\ref{#1}(#2)}}

\newcommand{\cale}{\mathcal{E}}
\newcommand{\calh}{\mathcal{H}}
\newcommand{\densities}{\mathcal{D}}
\newcommand{\divj}{{\textrm{div}\,\bm{J}}}

\usepackage[normalem]{ulem}

\usepackage{xcolor}
\definecolor{pastelgray}{rgb}{0.81, 0.81, 0.77}
\definecolor{beaublue}{rgb}{0.9, 0.9, 0.93}

\begin{document}
\title{Emulating Density Functional Theory Calculations via Empirical Interpolation}

\author{Daniel Lay\orcidlink{0000-0001-8947-391X}}
\affiliation{Physics Division, Argonne National Laboratory, Argonne, Illinois 60439-4801, USA}

\author{Pablo Giuliani\orcidlink{0000-0002-8145-0745}}
\affiliation{FRIB Laboratory, Michigan State University, East Lansing, Michigan 48824, USA}

\author{Kyle Godbey\orcidlink{0000-0003-0622-3646}}
\affiliation{FRIB Laboratory, Michigan State University, East Lansing, Michigan 48824, USA}

\begin{abstract}
\edef\oldrightskip{\the\rightskip}
\begin{description}
\rightskip\oldrightskip\relax
\setlength{\parskip}{0pt}

\item[Background] Nuclear density functional theory (DFT) is a suitable tool for predicting nuclear ground-state and fission properties. Statistical uncertainty quantification is desirable to make those predictions reliable, especially for nuclei far from stability. However, the computational cost associated with describing deformed nuclei in DFT makes such uncertainty quantification a challenge. In many solvers, the main computational bottleneck is the transformation of the wavefunction-dependent operators from coordinate to configuration space.

\item[Purpose] We explore the use of the empirical interpolation method (EIM) to speed up the coordinate-configuration transformations, effectively constructing DFT emulators for ground-state and fission properties. To train and test the emulator we vary the model parameters across their realistic posterior distribution. We consider both a simplified one-dimensional model, and realistic axially-deformed nuclei.

\item[Methods] We use Hartree-Fock-Bogoliubov (HFB) calculations to compute local densities and currents while varying the DFT model parameters. We consider sample nuclei from across the chart, from $A=60$ up to $A=254$, as well as a highly-deformed fission isomer. We construct one emulator for each case, and study the binding energy, quadrupole deformation, and excitation energy of the fission isomer.

\item[Results] In all nuclei, for all observables considered, the EIM emulator agrees with the DFT value to the precision of the original DFT calculations, using as few as 100 HFB calculations to build the emulator. For a given nuclear ground state or isomer, the emulator is able to predict all observables simultaneously. In terms of computational costs, the emulator provides an order-of-magnitude speedup over the original solver.

\item[Conclusions] EIM is a suitable emulation scheme for DFT, especially when high precision is desired as in model calibration and fission. It is also fairly straightforward to implement in existing HFB solvers. Thus, the EIM helps make statistical uncertainty feasible, improving the reliability of future predictions.

\end{description}
\end{abstract}

\date{\today}
\maketitle

\section{Introduction}

The quantum many-body is well known for exponentially increasing computational complexity with increasing number of particles. Density functional theory (DFT) scales polynomially, and as a consequence is an efficient method for approximately solving the many-body problem, especially for many-particle systems. In nuclear physics, DFT calculations are based on an effective nucleon-nucleon interaction whose model parameters are calibrated to experimental data~\cite{Schunck2019}. Uncertainties in the calibration must be propagated in order to make reliable predictions~\cite{Schunck2015jpg,Schunck2015epja,Mcdonnell2015}. Doing so in a Bayesian manner requires tens- or hundreds-of-thousands of model calculations~\cite{Phillips2021}, which is a significant computational burden. As such, model emulators are of great interest.

Various DFT emulation techniques have been developed, including neural networks~\cite{Lasseri2020,Lay2024b} and Gaussian processes~\cite{Verriere2022} for constrained calculations. An alternative approach, using projection-based emulators~\cite{Duguet2024}, such as the Reduced Basis Method (RBM)~\cite{Quarteroni2015,Hesthaven2016}, has been applied to unconstrained DFT calculations~\cite{Bonilla2022,Giuliani2023} and multireference calculations of excited states~\cite{Zhang2025}. Projection-based emulators require little training data and provide orders-of-magnitude speedup. Further, the emulation errors can be systematically reduced to any desired precision. Statistical uncertainties in DFT calculations for the binding energy of near-dripline nuclei~\cite{Erler2012,Neufcourt2020} and highly-deformed fission barriers~\cite{Mcdonnell2015,Schunck2015epja,Agbemava2017} vary between 0.5-3 MeV. Thus, high-precision emulators are desirable, and projection-based emulators are attractive for DFT studies.

Thus far, projection-based emulators for ground-state calculations have been restricted to spherical nuclei. Axial deformation, however, is critical for describing ground-state properties of nuclei~\cite{Bohr1998}, and fission studies~\cite{Schunck2016}. Thus, in this work, we develop an emulator based on the singular value decomposition (SVD) that targets the main bottleneck in many axially-deformed DFT calculations: the transformation from coordinate space to configuration space. This approach is both useful for, and straightforward to implement in, any DFT solver that makes this transformation. We apply our scheme to multiple case studies, and demonstrate that both the speedup and precision seen in projection-based emulators hold for ground-state and fission isomer calculations. 

This paper is organized as follows. First, Sec.~\ref{sec:dft} briefly reviews density functional theory, emphasizing the target of the emulator employed in this work. Next, Sec.~\ref{sec:toy-model} describes and applies the emulator to a simplified Skyrme-type energy density functional (EDF) in the spherically-symmetric case. Third, Sec.~\ref{section:skyrme-case} generalizes and applies the approach to a realistic Skyrme EDF, with axial deformation. Finally, in Sec.~\ref{Sec: conclusions} we present our conclusions and outlooks.

\section{Density Functional Theory}
\label{sec:dft}

We consider nuclear density functional theory (DFT) at the Hartree-Fock-Boguliubov (HFB) level, which describes quasiparticles moving independently in a self-consistent mean field. In this section we briefly review the relevant aspects of HFB theory; for a detailed description, see Ref.~\cite{Schunck2019} and references therein. 

Modern approaches begin by writing the total energy of the system, the energy density functional (EDF), as a sum of the particle-hole (ph) and particle-particle (pp) energy densities $\cale_\textrm{ph}$ and $\cale_\textrm{pp}$,
\begin{align}
	E[\rho(\bm{r}),\tilde{\rho}(\bm{r})]=\int d^3r\,\Big(\cale_\textrm{ph}(\bm{r})+\cale_\textrm{pp}(\bm{r})\Big).
\end{align}
where $\cale_\textrm{ph}$ is a functional of the one-body density $\rho_q$, the kinetic energy density $\tau_q$, and the spin-current density $J_{\mu\nu,q}$, while $\cale_\textrm{pp}$ depends on both $\rho_q$ and the pairing density $\tilde{\rho}_q$. The index $q=n,p$ describes the nucleonic species. We consider even-even nuclei, and so neglect time-odd densities and currents. Explicit expressions for the densities in terms of the Boguliubov transformation may be found in Ref.~\cite{Bender2003}.

The HFB equations follow from the variational principle, constrained such that the Boguliubov transformation is unitary, yielding the self-consistent eigenvalue problem~\cite{Schunck2016}
\begin{align}\label{eq:HFBMatrix}
    \mathcal{H}[\rho,\tilde{\rho}]
    \begin{pmatrix}
        U \\ V
    \end{pmatrix}_\mu
    =E_\mu
    \begin{pmatrix}
        U \\ V
    \end{pmatrix}_\mu.
\end{align}
$\mathcal{H}$ is the HFB matrix, $E_\mu$ is the quasiparticle energy of the $\mu$-th eigenstate, and $U,V$ are the matrices describing the Boguliubov transformation. Average particle number is fixed via the method of Lagrange multipliers. Constrained calculations enforce additional constraints, such as those on multipole moments (e.g. in potential energy surfaces (PESs)~\cite{Berger1984,Flynn2022,Schunck2016}). The discussion below is largely the same for constrained and unconstrained calculations, and is presented for the unconstrained case. We return to constrained calculations in the conclusion.

It is convenient to write $\calh$ in a block form, as
\begin{align}\label{eqn:hfb_block_mat}
    \calh=
    \begin{pmatrix}
        h & \tilde{h} \\ \tilde{h} & -h
    \end{pmatrix},
\end{align}
where the mean-field potential and pairing field are defined as~\cite{Dobaczewski1984}
\begin{align}
    h_{\mu\nu}=\frac{\delta E}{\delta\rho_{\nu\mu}},\quad \tilde{h}_{\mu\nu}=\frac{\delta E}{\delta\tilde{\rho}_{\nu\mu}}.
\end{align}
As is commonly done (e.g. in the UNEDF series of Skyrme parameterizations~\cite{Kortelainen2010,Kortelainen2012,Kortelainen2014}), we neglect isoscalar pairing, so eigenstates have definite nucleonic species.

To solve the HFB equations, we iteratively diagonalize $\calh$, as in Refs.~\cite{Vautherin1972,Vautherin1973,Baran2008,Schunck2016} and references therein. The diagonalization is efficient when carried out in configuration space, which is achieved by the transformation
\begin{subequations}
\begin{align}
    U_{q\mu}(\bm{r}\sigma)&=\sum_\nu U_{q,\mu \nu}\Phi_\nu(\bm{r}\sigma),\\
    V_{q\mu}(\bm{r}\sigma)&=\sum_\nu V_{q,\mu \nu}\Phi_\nu(\bm{r}\sigma).
\end{align}
\end{subequations}
Here, $\sigma$ is the usual spin index, and  $\{\Phi_\nu(\bm{r}\sigma)\}$ are the basis functions. This is the approach used in the axial codes HFBTHO~\cite{Perez2017} and DIRHB~\cite{Niksic2014}, as well as the triaxial code HFODD~\cite{Schunck2012}. The EDFs are defined in position space, so the transformation amounts to carrying out integrals of the form 
\begin{align}
    h_{\mu\nu}=\int d^3r\,d^3r'\,\sum_{\sigma\sigma'}\Phi_\mu^*(\bm{r}\sigma)h(\bm{r}\sigma,\bm{r}'\sigma')\Phi_\nu(\bm{r}'\sigma')
\end{align}
at each iteration, where $h(\bm{r}\sigma,\bm{r}'\sigma')$ is the mean field in position space. Analogous integrals are necessary for $\tilde{h}_{\mu\nu}$. Typical EDFs only depend on the local mean- and pairing fields $h(\bm{r},\sigma\sigma')$ and $\tilde{h}(\bm{r},\sigma\sigma')$~\cite{Schunck2019}.

\begin{figure}
    \centering
    \includegraphics[width=\linewidth]{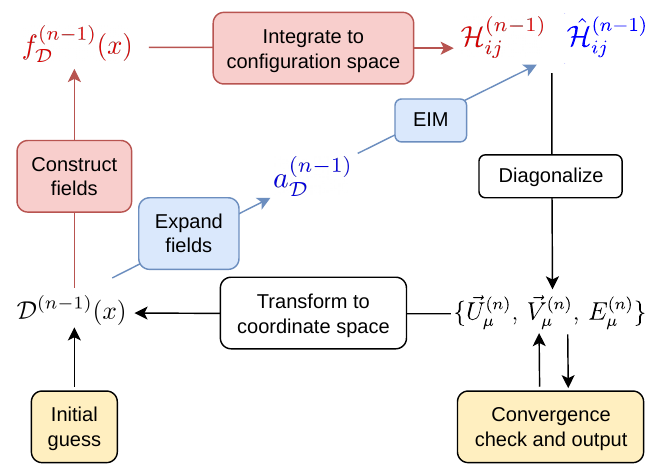}
    \caption{An illustration of the algorithm for a configuration-space self-consistent framework. The $\densities$ are the local densities and currents from Eq.~(\ref{eqn:hfb_densities}), each with a corresponding field $f_{\densities}$. The red boxes and arrows highlight the computational bottleneck, transforming the fields to configuration space~\eqref{eqn:field_transformations}. The blue arrows and boxes shows the EIM method, in which the fields are expanded in a precomputed basis with expansion coefficients $a_\densities$ at each iteration, leading to an approximate HFB matrix $\hat{\mathcal{H}}$ in configuration space. All variables in the diagram depend on the EDF parameters $\alpha$.}
    \label{fig:solver_diagram}
\end{figure}

This transformation is conveniently evaluated separately for the various local densities and currents
\begin{subequations}\label{eqn:hfb_densities}
\begin{align}
	\densities&\in \{\rho_q,\Delta\rho_q,\tau_q,\nabla\cdot\bm{J}_q,\ldots\}_{q=n,p}, \\
    \tilde{\densities}&\in \{\tilde{\rho}_q,\ldots\}_{q=n,p}.
\end{align}
\end{subequations}
For each, the contribution to the HFB matrix~\eqref{eqn:hfb_block_mat} can be written as
\begin{subequations}\label{eqn:field_transformations}
\begin{align}
	h^{\densities}_{\mu\nu}&=\int d^3r\,\sum_{\sigma\sigma'}g_{\mu\nu}^{\densities}(\bm{r},\sigma\sigma')\frac{\delta E_\textrm{HFB}}{\delta \densities}(\bm{r},\sigma\sigma'),\\
    \tilde{h}^{\tilde{\densities}}_{\mu\nu}&=\int d^3r\,\sum_{\sigma\sigma'}g_{\mu\nu}^{\tilde{\densities}}(\bm{r},\sigma\sigma')\frac{\delta E_\textrm{HFB}}{\delta \tilde{\densities}}(\bm{r},\sigma\sigma')
\end{align}
\end{subequations}
for particular functions $g$. The variations $f_{\densities}\equiv \dfrac{\delta E_\textrm{HFB}}{\delta \densities}$ are referred to as fields~\cite{Stoitsov2005}. For the particle densities, the fields are related to the central and spin-orbit potentials, and the effective mass~\cite{Vautherin1973}. It is, however, convenient to separate the $\rho_q$ and $\Delta\rho_q$ terms, as the functions $g_{\alpha\beta}^{\rho}$ and $g_{\alpha\beta}^{\Delta\rho}$ differ. Then, the full HFB matrix is a sum of these transformations:
\begin{align}
	h=\sum_{\densities}h^{\densities},\quad \tilde{h}=\sum_{\tilde{\densities}}h^{\tilde{\densities}}.
\end{align}
The self-consistent loop is shown in Fig.~\ref{fig:solver_diagram}. Note that, because the fields are defined in coordinate space, the inverse transformation is also required. In this work we consider axially-symmetric systems and take $\{\Phi_\nu\}$ to be deformed harmonic oscillator (HO) states, for which explicit expressions may be found in Refs.~\cite{Vautherin1973,Stoitsov2005}. Then, the HFB matrix is block-diagonal, with each block computationally inexpensive to diagonalize, and the primary computational bottleneck is the transformation from coordinate space to configuration space, shown as red arrows and boxes in Fig.~\ref{fig:solver_diagram}.

EDFs often contain non-affine parameter dependence, i.e. $\mathcal{H}$ is not an EDF parameter times a coordinate-dependent function. For instance, the Skyrme~\cite{Kohler1976} and Gogny~\cite{Robledo2019} EDFs contain a term of the form $(\rho_p+\rho_n)^\gamma$ with $\gamma$ a tunable parameter, descending originally from a contact three-body interaction~\cite{Vautherin1972}. Additionally, non-affine terms are present in more recently-developed EDFs, such as the Fayans~\cite{Fayans1998,Reinhard2017} and GUDE~\cite{Zurek2024} EDFs. Thus, the transformation to configuration space must be carried out at each iteration, and emulation techniques such as the reduced basis method~\cite{Bonilla2022} can become less efficient. 

The integrals that transform to configuration space are therefore the target of the emulator developed in this work. For illustrative purposes, we consider next a simple case.

\section{A Simple Case}
\label{sec:toy-model}

The Gross-Pitaevskii (GP) equation was introduced originally to study superfluidity in bosonic systems~\cite{Gross1961,Pitaevskii1961}. The one-dimensional GP equation has been used as a test case for a number of dimensionality reduction techniques in both bosonic~\cite{Pichi2020} and fermionic~\cite{Bonilla2022,Bakurov2025} systems. For fermionic systems, we consider the modified Hamiltonian
\begin{align}
    H_\textrm{GP}=-\frac{d^2}{dx^2}+kx^2+q\rho^\gamma(x),
\end{align}
where the density $\rho$ is defined as
\begin{align}
    \rho(x)=\sum_{i=1}^N|\psi_i(x)|^2,
\end{align}
with $\psi_i$ the $i$-th eigenstate and $N$ the number of particles considered. This Hamiltonian is chosen precisely for the non-affine parameter dependence $\rho^\gamma$. The goal is to efficiently solve for the lowest $N$ eigenstates of ${H}_\textrm{GP}$ as the parameters $\boldsymbol{\alpha}=(k,q,\gamma)$ are varied. 

\subsection{Empirical Interpolation}
\label{subsec:gp-ei}
As in the HFB case, we diagonalize the GP Hamiltonian in configuration space, defined by the basis $\{\phi_j(x)\}_{j=1}^M$. The transformation is
\begin{align}
    \langle\phi_j|{H}_\textrm{GP}|\phi_k\rangle&=T_{jk}+kV_{jk}+q\langle\phi_j|\rho^\gamma|\phi_k\rangle,
\end{align}
where
\begin{align}
    T_{jk}=\bigg\langle\phi_j\bigg|-\frac{d^2}{dx^2}\bigg|\phi_k\bigg\rangle\quad \textrm{and}\quad
    V_{jk}=\langle\phi_j|x^2|\phi_k\rangle
\end{align}
are the transformation of the kinetic and harmonic potentials, respectively. Note that $T$ and $V$ can be computed once, independent of both the iteration and the model parameters $\bm{\alpha}$. Conversely, the self-consistent potential must be recomputed in both cases.

The approach we follow to mitigate this issue is the Empirical Interpolation Method (EIM). EIM was introduced in Refs.~\cite{Barrault2004,Chaturantabut2010}, its convergence properties analyzed in Ref.~\cite{Maday2016}, incorporated in recent textbooks~\cite{Quarteroni2015,Benner2017,Brunton2019}, and used in nuclear theory in Refs.~\cite{Anderson2024,Odell2024,Maldonado2025,catacora2025wavefunction}. For a function $f(\boldsymbol{\alpha};x)$ that depends on both model parameters $\boldsymbol{\alpha}$, and a coordinate $x$, the EIM constructs the approximation
\begin{align}\label{eqn:ei}   
f(\boldsymbol{\alpha};x)\approx \hat f(\boldsymbol{\alpha};x)=\sum_{i=0}^{N_\textrm{EIM}}a_i(\boldsymbol{\alpha})\hat{f}^{(i)}(x),
\end{align}
for a set of basis functions $\{\hat{f}^{(i)}(x)\}$ and expansion coefficients $\{a_i(\boldsymbol{\alpha})\}_{i=0}^{N_{\textrm{EIM}}}$.

To determine $\hat{f}^{(i)}(x)$, we evaluate $f(\alpha;x)$ for a number of parameters $\alpha_1,\ldots,\alpha_m$ (the ``training'' dataset), then form the matrix 
\begin{align}
    \mathcal{F}_{ij}\equiv f(\boldsymbol{\alpha}_i;x_j).
\end{align}
The $x_j$ are the coordinate mesh points on which $\rho$ is defined (typically numerical quadrature points used efficiently evaluate evaluate the integral transformations~\cite{Stoitsov2005}). The singular value decomposition (SVD)~\cite{Brunton2019} of $\mathcal{F}$ yields an orthonormal basis informed by the exact evaluations, ordered from most to least important by the singular values $\Sigma_i$. The expansion coefficients $a_i(\alpha)$ are then also ordered by importance for all $\alpha$ within some range of the sample solutions. In practice, this range is often fairly large~\cite{Bonilla2022}. We choose some threshold $\varepsilon_\textrm{EIM}$, and only include basis functions where $\Sigma_i/\Sigma_0\geq\varepsilon_\textrm{EIM}$. This corresponds to an error in the basis truncation of order $\mathcal{O}(\varepsilon_\textrm{EIM})$; as will be demonstrated the error in the final result is typically of the same order.

With this affine decomposition, the matrices
\begin{align}
    \hat{f}_{jk}^{(i)}=\langle\phi_j|\hat{f}^{(i)}|\phi_k\rangle
\end{align}
can be precomputed. Thus, at each iteration, the transformation to configuration space only requires determining the expansion coefficients $a_i(\bm{\alpha})$. This is the step shown in blue in Fig.~\ref{fig:solver_diagram}. Note that, for the GP case, there is only one density to consider, $\densities=\rho$.

The expansion coefficients are determined by evaluating both sides of Eq.~(\ref{eqn:ei}) at some points $\{x_k\}$, called the collocation points. This results in the (possibly overdetermined) matrix equation
\begin{align}
    A\bm{a}=\bm{y},\quad\textrm{where}\quad A_{ji}\equiv \hat{f}^{(i)}(x_j),\quad y_j=f(\bm{\alpha};x_j).
\end{align}
In general $i$ and $j$ have different ranges; the number of collocation points is typically at least as large as $N_{\textrm{EIM}}$. The optimal least-squares solution is the Moore-Penrose pseudoinverse, $\bm{a}=(A^TA)^{-1}A^T\bm{y}$. The matrix $(A^TA)^{-1}A^T$ can precomputed and stored. Here, we take the collocation points to be the original grid.

EIM as phrased above requires being able to evaluate $f(\bm{\alpha};x)$ at arbitrary $(\bm{\alpha},x)$. This cannot be done for the density-dependent term in ${H}_{\textrm{GP}}$, $q\rho^\gamma$. Instead, the expansion coefficients are updated iteration-by-iteration: given coefficients $\bm{a}^{(n)}$ at iteration $n$, we compute and diagonalize
\begin{align}
    \hat{H}_{\textrm{GP},jk}^{(n)}\approx T_{jk}+kV_{jk}+\sum_{i=0}^{N_\textrm{EIM}}a_i^{(n)}(\alpha)\hat{f}^{(i)}_{jk}.
\end{align}
The resulting eigenstates $\psi_i^{(n+1)}(x)$ are then used to compute $\rho^{(n+1)}$, and thereby $q[\rho^{(n+1)}]^\gamma$. This takes the place of the exact potential $f(\bm{\alpha};x_j)$, and is used to determine $\bm{a}^{(n+1)}$.

We repeat this iterative scheme until convergence is achieved, defined as 
\begin{align}
    \max |\bm{a}^{(n+1)}-\bm{a}^{(n)}|\leq \varepsilon_{\textrm{conv}}
\end{align}
for a chosen tolerance $\varepsilon_{\textrm{conv}}$. Convergence is greatly improved when using a mixing scheme (i.e. using a particular linear combination of $\bm{a}^{(1)},\ldots,\bm{a}^{(n)}$ in place of $\bm{a}^{(n)}$); in this section we use a linear mixing scheme~\cite{Baran2008}.

\subsection{The Reduced Basis Method}\label{subsec:toy-rbm}
The diagonalization of $H_{\textrm{GP}}$ is typically the next most computationally expensive step. Here we briefly review the reduced basis method (RBM) as a tool for achieving further run time improvements; see the recent review~\cite{Duguet2024} for further details.

The RBM is simply a basis expansion approach, in a tailored basis. It has been applied to self-consistent solvers~\cite{Bonilla2022,Giuliani2023} and widely applied in other areas of nuclear theory~\cite{Duguet2024}. When few wavefunctions are required, such as in one-particle or spherical coordinate systems, fewer than 10 basis wavefunctions are required to achieve sub-percent errors~\cite{Bonilla2022}. As a consequence, in self-consistent cases, considerable speedups have been achieved despite transforming to configuration space at each iteration~\cite{Bonilla2022,Giuliani2023}.

Analogous to the expansion functions $\{\hat{f}^{(i)}(x)\}$, the basis wave functions are determined using the SVD of the wave functions as $\alpha$ is varied. Note that this introduces another threshold $\varepsilon_\textrm{RBM}$ that must be chosen to truncate the SVD. 

When multiple eigenstates are needed, one may either construct a basis for each wave function separately, or construct a single basis that describes all wave functions simultaneously. The former requires grouping similar wavefunctions together, which in many cases is enabled by the Courant nodal domain theorem~\cite{Courant1985} - for instance, in 1D, the theorem guarantees that the eigenstate corresponding to the lowest eigenvalue has zero nodes. Conversely, when this theorem does not hold, the basis must be constructed for all wave functions simultaneously. In the realistic case, Sec.~\ref{section:skyrme-case}, the latter is the relevant case. So, we only consider the latter case here.

In this section, we consider EIM both using the HO basis and the RBM. 

\subsection{Simple Case Results}
We first compare three different traditional approaches: an HO basis expansion, a uniform finite-difference grid discretization, and a Chebyshev~\cite{Boyd2001} grid discretization. We compare two scenarios: one with $N=5$ particles, and one with $N=20$. To select our parameters $\bm{\alpha}$, we fix a mean value of $\bm{\alpha}=(1,5,5)$, and sample 1000 points from a normal distribution with variance of 15\% of the mean value; this is reflective of the spread in the HFB case discussed in Sec.~\ref{section:skyrme-case}.

In the HO expansion, the number of oscillator shells is varied from $20-150$. The grid discretization schemes increase the number of gridpoints used, between $100-1000$ and $20-200$ for finite difference and Chebyshev grids, respectively. The reference (``exact'') eigenvalue is taken from the highest-precision calculation, which is in agreement between the three methods, and each solver uses the same convergence tolerance $\varepsilon_\textrm{conv}=10^{-8}$. All timing calculations are performed on one core on an AMD Ryzen 5 7640U processor, with other processes halted. The initial guess is the lowest-lying eigenstates of ${H}_{\textrm{GP}}$ with $q=0$. 

\begin{figure}[h]
    \centering
    \includegraphics[width=\linewidth]{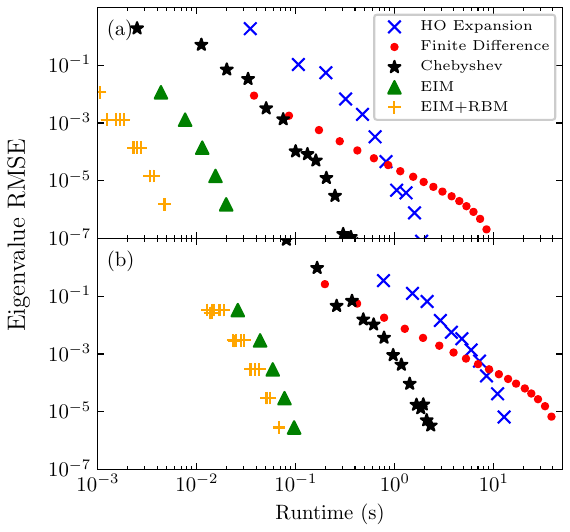}
    \caption{The root-mean-square error on the lowest $N$ eigenvalues versus the total runtime of the self-consistent solver for the GP equation. Panel (a) is $N=5$ particles; panel (b) is $N=20$. Both the RMSE and the run time are averaged over the different $\bm{\alpha}$ values.}
    \label{fig:gp-cat}
\end{figure}

The timing results for the exact solvers are shown in Fig.~\ref{fig:gp-cat}. For each solver, as the resolution is increased (by including more basis states, or decreasing the mesh spacing), the error on the eigenvalues decreases. Conversely, the total runtime increases.

The HO expansion option is slower than the finite difference solver, up to a crossover runtime of about 1 second for $N=5$. Additionally, the discretized grid based on Chebyshev meshpoints is faster than the finite difference solver at low accuracy, and scales similar to the HO expansion at high precision. The difference between the Chebyshev and HO expansion runtimes is due to the transformation to configuration space. For a given solver, the runtime difference with increasing resolution is due to the increased matrix size, not the total number of iterations, since the same tolerance is used for each resolution.

Next, we consider the EIM, reconstructing the wave functions on the original grid. We take as training 100 randomly-selected parameter sets from the 1000 samples generated previously, and use the remaining 900 as validation. These samples are fixed between EIM variants. Figure~\ref{fig:gp-cat} also shows the runtime and error for the validation data set. As EIM is an SVD-based emulator, the error on the training dataset is typically of order $\mathcal{O}(\varepsilon_{\textrm{EIM}})$; this estimation does not require running the emulator for training points. Nevertheless, we have checked, and performance on the training dataset (not shown) is similar.

For representing the wave functions, we consider both an HO and a tailored (reduced) basis. The tailored basis was constructed for all wave functions simultaneously. A minor speedup is achieved when building a basis for positive and negative parity states independently (not shown). The tolerances $\varepsilon_\textrm{EIM},\,\varepsilon_\textrm{RBM}$ are varied between $10^{-3}-10^{-7}$, which corresponds to varying the number of basis functions. In the self-consistent solver, eventually the difference iteration-to-iteration is below the truncation error from the expansion(s) (see Fig.~\ref{fig:error-vs-basis-size} below); thus, a smaller $\varepsilon_{\textrm{conv}}$ does not automatically lead to a more accurate solution. As a heuristic measure, in this section we set $\varepsilon_\textrm{conv}=10\times\textrm{max}\,(\varepsilon_\textrm{EIM},\varepsilon_\textrm{RBM})$. 

Consider Fig.~\ref{fig:gp-cat}(a) for the system with $N=5$ wavefunctions. EIM in the HO basis is consistently 10 times faster than the Chebyshev approach, and 100 times faster than the HO solver, for equivalent precision. Despite the different solver tolerances, the speedup is almost entirely because of the reduced runtime per iteration. That EIM is only one order of magnitude faster than the Chebyshev grid suggests that pseudospectral methods on tailored grids may provide an additional speedup. The EIM $+$ RBM approach demonstrates that using a tailored basis speeds up the runtime by roughly a factor of 5, with similar precision.

Figure~\ref{fig:gp-cat}(b) shows that most of the conclusions hold for $N=20$. The notable exception is that the speedup from the tailored wave function basis is drastically decreased. This is because the wave function basis size must be enlarged to capture higher-lying states, and thus using a tailored basis does not drastically reduce the size of the matrix that is being diagonalized. This suggests that a global wave function basis becomes less useful as more states are desired; we return to this point in the next section.

We conclude that the EIM is a useful emulation scheme for simple self-consistent problems. Next, we examine the extent to which this is true for a realistic Skyrme case.

\section{Skyrme HFB Theory}\label{section:skyrme-case}

In this section, we first generalize the EIM approach to HFB theory using the realistic Skyrme EDF, allowing for axial deformation. We then discuss the accuracy, speedup, and extrapolation capabilities, and conclude by applying the method to a fission isomer example.

\begin{figure*}
    \centering
    \includegraphics[width=\linewidth]{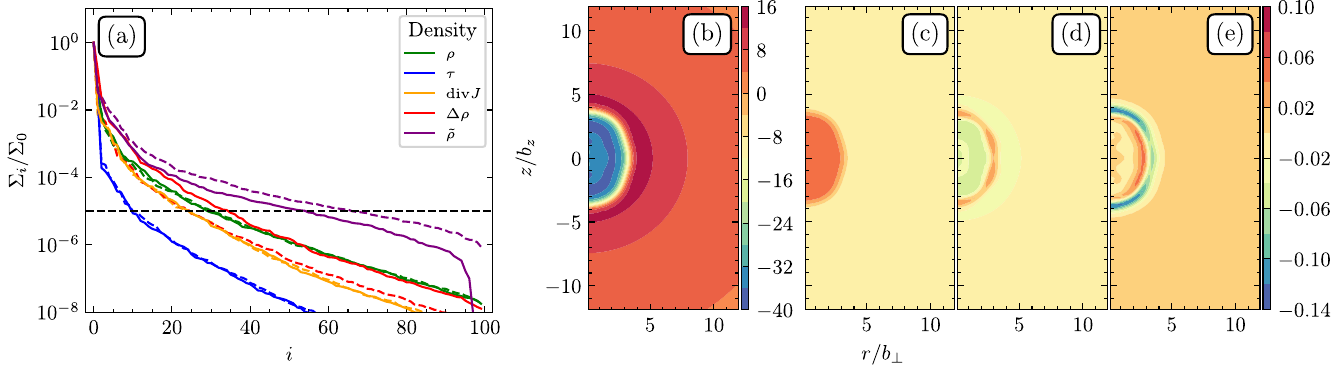}
    \caption{(a): The singular values for the fields for the ground state of $^{254}$Fm, using 25 oscillator shells at the HFB level. Proton (neutron) fields are shown as solid (dashed) lines. An example threshold of $\varepsilon_\textrm{EIM}=10^{-5}$ is shown as a black horizontal dashed line. (b): The $\rho_p$ field for a representative EDF parameter, in MeV-fm${}^{3}$. (c)-(e): The first three basis fields from the SVD of the $\rho_p$ field. The sample field is unnormalized, while each basis field (defined on a finite mesh) sums to 1. The axes in (b)-(e) are scaled by the oscillator widths $b_z$ and $b_\perp$ used in HFBTHO~\cite{Perez2017} and this work. See Sec.~\ref{subsubsec:accuracy} for discussion on the sample points used for the SVD for this case.}
    \label{fig:skyrme-fields-svd-combined}
\end{figure*}

\subsection{The Approach}\label{subsec:skyrme-differences}
While the approach from the toy model largely carries over to the axial HFB case, there are three important differences, the first concerning the EIM. We apply EIM to each field separately, as each consists of a different integral, recall Eq.~(\ref{eqn:field_transformations}). However, the decay rate of the singular values varies between the fields, as shown in Fig.~\ref{fig:skyrme-fields-svd-combined}(a) for a representative nucleus. The singular values for the kinetic field decay quickest, due to the constant value of $-\hbar^2/(2m_q)$ outside of the localized nucleus, originating from the usual single-particle kinetic energy density $\mathcal{E}_{\textrm{kin}}=\hbar^2\sum_q\tau_q/(2m_q)$~\cite{Vautherin1972}. Those of the pairing field $\tilde{\rho}_q$ decay rather slowly, reflecting the subtle structural changes with varying EDF parameters. Thus, the total number of HFB samples required to construct the emulator is determined by the pairing density $\tilde{\rho}_q$. 

Panel (b) of Fig.~\ref{fig:skyrme-fields-svd-combined} shows the proton field for one set of EDF parameters. It is mostly localized, with a nonzero tail due to the Coulomb potential. The first three basis fields, panels (c)-(e), are localized in the same spatial region as the sample. They capture first the near-saturation of the nuclear interior (Panel~(c)), then variations around the nuclear surface and fine details on the interior (Panels~(d)-(e)). The basis fields for the other fields behave analogously.

As will be seen shortly, the choice of collocation points is now important. We compare using the original grid to using the MaxVol algorithm to determine $N_\textrm{EIM}$ collocation points that maximize the information contained in the sub-matrix $A_{ji}$, $0\leq i,j\leq N_\textrm{EIM}$; see Refs.~\cite{Olshevsky2010,Odell2024} for more details on the algorithm. Since the basis fields are localized, the MaxVol algorithm selects collocation points inside the localized nuclear density. However, these collocation points are typically different for each field. The fields typically depend on multiple densities (e.g. $f_{\rho_q}$ depends on $\tau,\,\Delta\rho$, and $\divj$), and therefore the various densities must be evaluated on a common mesh; we take this to be the set-wise union of the collocation points. One iteration of the self-consistent solver updates the expansion coefficients for each field, $\bm{a}_{\densities}(\alpha)$ and $\bm{a}_{\tilde{\densities}}(\alpha)$. Otherwise, the iterative scheme follows that of Sec.~\ref{subsec:gp-ei}, using Broyden mixing~\cite{Baran2008}; see also the blue portion of Fig.~\ref{fig:solver_diagram}.

The second difference from the problem discussed in Sec.~\ref{sec:toy-model} concerns the quasiparticle wave functions. In the toy model (and the Hartree-Fock case), wavefunctions either contribute or do not. That is, to describe a nucleus with $Z$ protons, exactly $Z$ proton single-particle wave functions are required. In HFB theory, states have partial occupations, based on the lower component of the quasiparticle wavefunction: the occupation
\begin{align}
    N_{q\mu}=\int d^3r\,\sum_\sigma |V_{q\mu}(\bm{r}\sigma)|^2
\end{align}
becomes a continuous number between 0 and 1. Thus, the total number of wavefunctions that needs to be tracked is much larger than the number of nucleons in the system.

Further, for EDFs with a zero-range pairing term, the pairing strength must be regularized, otherwise the density matrix diverges with increasing quasiparticle energy~\cite{Dobaczewski1984,Dobaczewski1996,Lalit2026}. As in the HFBTHO code~\cite{Stoitsov2005}, we handle this divergence using a cutoff regulator, i.e. we only consider states with quasiparticle energy $E_\mu\leq 60$ MeV. Now, the number of states that must be captured varies with the EDF parameters, and additionally, some states cross the 60 MeV threshold as the EDF parameters vary (meaning they are excluded in the SVD). Therefore, we include additional states above the cutoff threshold in the RBM construction, and indeed, if too few states are included, the accuracy of the emulator is severely degraded.

Moreover, individual wave functions cannot be tracked as the EDF parameters are changed, as level crossings lead to wave functions occupying a nearly-degenerate subspace. Strictly speaking, wave functions can be tracked through the crossings; this can be seen even in two-level systems. However, the ubiquity of crossings drastically increases the number of required calculations, reducing the utility of the emulator. Thus, we build a reduced basis for all of the wave functions simultaneously.

The third difference concerns the densities and currents. Some terms in the EDF, such as the direct Coulomb field,
\begin{align}
    V_{\textrm{Coul}}^D[\rho](\bm{r})=e^2\int d^3r'\,\frac{\rho_p(\bm{r}')}{|\bm{r}-\bm{r}'|},
\end{align}
 depend on the density at every gridpoint. Similarly, when considering approximate particle number restoration with the Lipkin-Nogami (LN) approach~\cite{Lipkin1960,Nogami1964}, the pairing energy $E_{\textrm{pair},q}$ and average pairing gap $\bar{\Delta}_q$ are commonly computed as spatial integrals over $\rho_q$ and $\tilde{\rho}_q$~\cite{Stoitsov2003,Stoitsov2005}:
\begin{subequations}
    \begin{align}
        E_{\textrm{pair},q}&=-\frac{1}{2}\int d^3r\,\tilde{h}_q(\bm{r})\tilde{\rho}_q(\bm{r}),\\
        \bar{\Delta}_q&=\frac{1}{N_q}\int d^3r\,\tilde{h}_q(\bm{r})\rho_q(\bm{r}),
    \end{align}
\end{subequations}
where $N_q$ is the number of nucleons of species $q$. So, certain densities must be reconstructed at every gridpoint, even if the corresponding field does not. 

An additional speedup is technically possible by applying EIM to each density that would otherwise be reconstructed. As observed in Ref.~\cite{Lv2026} in the spherical case and in this work in the axially deformed case, the various densities and currents also admit low-dimensional representations computable via the SVD. However, both $\rho_p$ and $\tilde{\rho}_q$ are quick to reconstruct, and so the obtained speedup is quite small. Note also that, for these three quantities, one may instead write the integrals as traces in the HO basis, circumventing the reconstruction problem entirely. Nevertheless, it may sometimes be useful to apply EIM to some or all of the densities/currents, in addition to the fields.

\subsection{Results}

We use the Skyrme EDF, parameterized as in the UNEDF calibrations~\cite{Kortelainen2010,Kortelainen2012,Kortelainen2014}, and we neglect the tensor energy density as in the UNEDF0 and UNEDF1 calibrations~\cite{Kortelainen2010,Kortelainen2012}. As mentioned previously, the relevant feature for this work is the density-dependent coupling constant $C_t^{\rho\rho}=C_{t0}^{\rho\rho}+C_{tD}^{\rho\rho}\rho_0^\gamma$, as it is the impetus for the EIM emulation approach. 

To generate physically-relevant samples to test EIM, we took 500 samples from the posterior parameter distribution of the UNEDF1 calibration~\cite{Mcdonnell2015}. Thus, each test below varies all 12 EDF parameters simultaneously. We then used the code HFBTHO v3~\cite{Perez2017} to solve the unconstrained HFB equations for each sample parameter. Note that, for full consistency with the UNEDF1 calibration, the LN procedure should be used to approximately restores particle-number symmetry~\cite{Stoitsov2005} in all tests. For illustrative purposes, we demonstrate our emulator with and without the LN procedure.

All emulator results are implemented in Python. Below, we summarize the accuracy across the nuclear chart, discuss the resulting speedup, comment on the extrapolation capability of the EIM, and apply EIM to fission isomer properties.

\subsubsection{Accuracy}\label{subsubsec:accuracy}
To demonstrate the emulator performance across the nuclear chart, we first consider representative nuclei: $^{60}$Ni, $^{156}$Gd, $^{236}$Pu, and $^{254}$Fm. Depending on the EDF parameters, the ground state of $^{60}$Ni may be either spherical or deformed; for the remaining nuclei, the ground state is always (prolate) deformed. We use 20 oscillator shells for $^{60}$Ni and 25 for the rest. We use 100 randomly-selected EDF parameters to construct an emulator for each nucleus, which are approximately evenly distributed throughout the parameter region, and evaluate the emulator on the remaining 400 validation points. We vary the tolerance $\varepsilon_{\textrm{EIM}}\in[10^{-6},10^{-2}]$, which, per Fig.~\ref{fig:skyrme-fields-svd-combined}, corresponds to increasing the number of EIM basis functions. We diagonalize $\calh$ in the HO basis.

\begin{figure}
    \centering
    \includegraphics[width=\linewidth]{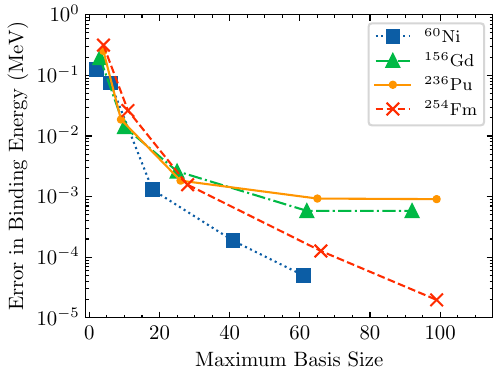}
    \caption{The average error on the binding energy, taken across the validation data set, versus the maximum number of EIM basis functions. $^{60}$Ni uses 20 oscillator shells; $^{156}$Gd, $^{236}$Pu, and $^{254}$Fm use 25. All cases are the unconstrained ground state; all but $^{254}$Fm are at the LN level. All errors are taken across the validation set.}
    \label{fig:error-vs-basis-size}
\end{figure}

Figure~\ref{fig:error-vs-basis-size} shows the emulator error in the binding energy plotted against the number of EIM basis functions on the validation dataset. For small basis sizes, the error decays exponentially, with a root-mean-square error of approximately 1 keV with as few as 20 basis fields (compare with errors of 100~keV achieved in Refs.~\cite{Verriere2022,Lay2024b,Lasseri2020}). For both $^{60}$Ni and $^{254}$Fm, the error continues to decrease to the largest basis size considered. For $^{156}$Gd and $^{236}$Pu the error plateaus because, eventually, $\varepsilon_\textrm{EIM}$ is below the convergence tolerance for the training data, so the emulator is trying to mimic noise. Were we to continue decreasing $\varepsilon_\textrm{EIM}$ for $^{60}$Ni and $^{254}$Fm, we expect the error to plateau as well.

We note that, for $^{60}$Ni with a small basis size, a few emulated runs do not converge. These runs are excluded from the plot. Nonconvergence is unique to $^{60}$Ni, and is not an issue for larger basis sizes. Note that convergence issues are somewhat common in self-consistent solvers; Broyden mixing greatly ameliorates these issues but does not entirely solve them. While the emulator fails to produce an answer, it does not produce a wrong answer, in contrast to neural network emulators~\cite{Lay2024b}. When in doubt, one may check if the solution is converged by performing one real self-consistent iteration as additional verification, though this strategy is not unique to the EIM approach~\cite{Verriere2022}. Thus, EIM is useful despite rare convergence issues.

To understand the error behavior, recall that solving the HFB equations is equivalent to minimizing the HFB energy. As such, the HFB energy in the vicinity of the true solution is quadratic in the density matrix~\cite{Ring1980,Blaizot1986}. Constraints, such as those on the average particle number, are satisfied by the emulated solution. Thus, an error $\delta$ in the densities corresponds to an error $\delta^2$ in the HFB energy. Indeed, for binding energy of the order of 1 GeV, a truncation error on the fields of $\sim10^{-3}$ corresponds roughly to a binding energy error of $\sim10$ keV, a relative error of $\sim10^{-5}$, approximately consistent with the expected quadratic behavior.

We conclude that the EIM can be made arbitrarily precise, given sufficient, and sufficiently precise, training data. In many cases - such as binding energy~\cite{Erler2012,Neufcourt2020,Buskirk2024} and fission~\cite{Schunck2016,Lay2024a} studies, as well as model calibration efforts~\cite{Kortelainen2010,Kortelainen2012,Kortelainen2014} - relative energies are important, and they vary on the MeV, rather than the GeV, scale. Thus, the high precision of this emulator is quite useful.

\subsubsection{Speedup}
Next, we discuss the resulting speedup by focusing on $^{254}$Fm. To check the robustness of EIM, we consider a different number of major oscillator shells. For each, we simultaneously vary $\varepsilon_\textrm{EIM},\varepsilon_\textrm{RBM}\in[10^{-6},10^{-2}]$, and consider both global and MaxVol reconstruction. We also vary the number of wave functions above the 60 MeV cutoff threshold that are included in the wave function SVD. Again, we take as training set 100 randomly-sampled EDF parameters, with the remaining 400 as validation. All timing is evaluated while forced to run on one core on an AMD Ryzen 5 7640U processor, with other processes halted. 

\begin{figure}
    \centering
    \includegraphics[width=\linewidth]{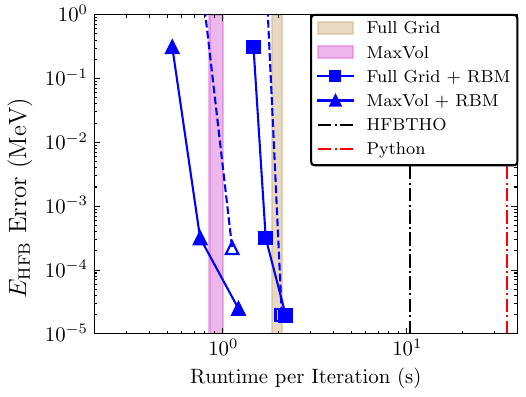}
    \caption{The runtime per iteration versus the root-mean-square error on the HFB energy for $^{254}$Fm at the HFB level, using 25 oscillator shells. The exact solvers are shown with vertical lines. All other data points use EIM, evaluated on either the full grid or the MaxVol collocation points as indicated. Points using the RBM are labeled as such. The solid bands indicate the range of run times while varying $\varepsilon_{\textrm{EIM}}\in[10^{-6},10^{-2}]$, for which the accuracy is shown in Fig.~\ref{fig:error-vs-basis-size}. The filled (open-face) symbols use all (zero) wave functions above the 60 MeV threshold in the RBM SVD, with $\varepsilon_{\textrm{EIM}}$ fixed at $10^{-5}$.}
    \label{fig:skyrme-cat-plot}
\end{figure}

First, we focus on the case using 25 major oscillator shells, for which the accuracy while varying $\varepsilon_{\textrm{EIM}}$ is shown in Fig.~\ref{fig:error-vs-basis-size}. Figure~\ref{fig:skyrme-cat-plot} shows the runtime per iteration of the different emulator configurations, as well as that of HFBTHO and an equivalent Python HFB solver. First, EIM with the HO basis and a global reconstruction provides a factor of 5 speedup over HFBTHO, and an order of magnitude speedup over the Python code. Most of the remaining computational time is spent diagonalizing the HFB matrix, in contrast with spherical systems~\cite{Bonilla2022,Giuliani2023}. So, the runtime per iteration is largely independent of $\varepsilon_{\textrm{EIM}}$; we represent this with the vertical band instead of individual points. The total number of iterations required to achieve convergence is largely unchanged.

A further speedup factor is obtained using the MaxVol reconstruction strategy, resulting in a net 10-times speedup over HFBTHO; otherwise, similar conclusions hold from the global reconstruction strategy.

Minor additional speedups are observed by using a tailored wavefunction basis (the RBM case) and reducing $\varepsilon_{\textrm{RBM}}$, as shown by the filled symbols in Fig.~\ref{fig:skyrme-cat-plot} with $\varepsilon_{\textrm{EIM}}=10^{-5}$. This is due to marginally smaller matrices that are being diagonalized, though the run time for the $10^{-5}$ threshold is actually longer than using the HO basis. As a tradeoff for improved runtime, the accuracy decreases. Decreasing the number of states included above the energy cutoff, as shown by open symbols, introduces MeV-scale errors, even for moderate $\varepsilon_{\textrm{RBM}}=10^{-3}$, as was mentioned in Sec.~\ref{subsec:skyrme-differences}. Similar conclusions hold for other thresholds $\varepsilon_{\textrm{EIM}}$, and for the two reconstruction strategies.

When considering 20 and 30 oscillator shells, the story is largely the same. For 20 (30) shells, the overall runtime per iteration is smaller (larger). The obtained speedup, meanwhile, is smaller (larger), as the transformation to the HO basis involves fewer (more) numerical integrals. Notably, the speedup using a tailored basis is larger (smaller) than the 25 shells case, as in the toy model, Fig.~\ref{fig:gp-cat}, and for the same reason. Due to both the potential for large error, and the somewhat reduced runtime improvement, we neglect the RBM case for the rest of the work.

EIM therefore provides a useful speedup over the original solver. However, it is still an iterative method, and hence slower than neural network-based emulation schemes (which often evaluate the densities in one call, rather than in a self-consistent loop~\cite{Verriere2022}). On the other hand, EIM is able to extrapolate in parameter space with some reliability, and predicts quantities that it is not ``trained'' on, as the output is not a number but the full HFB wavefunction and the assorted densities. The remainder of the work discusses these two points, and we return to the runtime discussion in the conclusion.

\begin{figure}
    \centering
    \includegraphics[width=\linewidth]{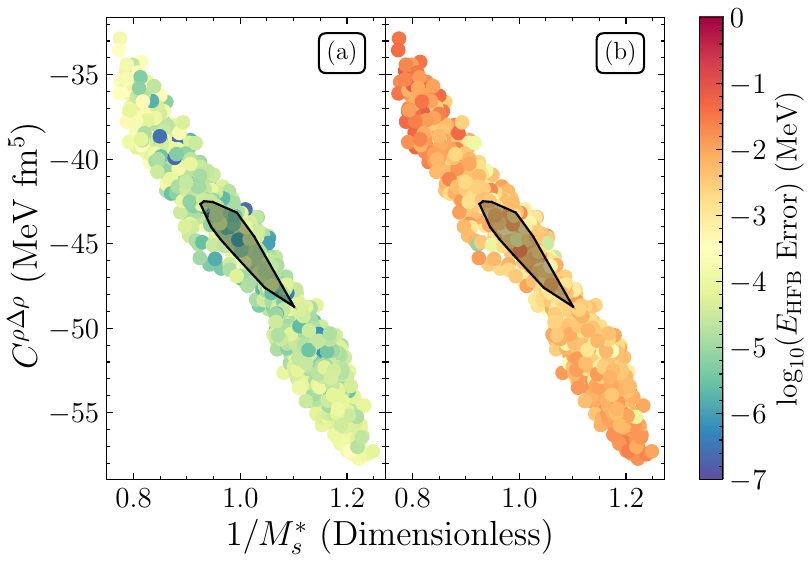}
    \caption{The error on $E_\textrm{HFB}$ for the unconstrained ground state of $^{254}$Fm, projected onto a representative subspace of the EDF parameters. (a) and (b) correspond to $\varepsilon_\textrm{EIM}=10^{-6}$ and $10^{-4}$, respectively. The black shaded region shows the convex hull of the training points in the $(1/M_s^*,C^{\rho\Delta\rho})$ space.}
    \label{fig:extrapolation}
\end{figure}

\subsubsection{Extrapolation}
As touched on in Sec.~\ref{subsec:gp-ei}, the EIM is able to extrapolate in parameter space. This is a fairly common feature of projection-based emulators~\cite{Bonilla2022,Frame2018}. This is in contrast to neural network schemes that tend to extrapolate poorly, even fairly close to the training dataset~\cite{Lay2024b}. In particular, the EIM is able to extrapolate within the posterior parameter region determined in the UNEDF1 calibration.

To demonstrate this, we compute the geometric mean parameter from the UNEDF1 posterior samples, and take as training data the 50 nearest neighbors, after normalizing the parameters to zero mean and unit variance. The projection of the parameters onto a representative subspace is shown in Fig.~\ref{fig:extrapolation}. The error on the binding energy is consistent with the previous sections, and does not degrade out to the edge of the parameter region.

This extrapolation ability is potentially useful for future EDF calibration efforts, especially for exploring high-dimensional parameter spaces. Further, projection-based emulators have been noted to converge even when the original solver does not, albeit in spherically-symmetric cases~\cite{Bonilla2022}. EIM may be understood as a projection-based emulator, in which the fields are projected on a low-dimensional manifold that describes the training data. So, EIM likewise may likewise enable exploration of larger parameter spaces. 

Extrapolation is also likely to be helpful when changing e.g. the number of nucleons, as when generating potential energy surfaces for large-scale fission studies~\cite{Lay2024b}, given a sufficiently expressive basis for the HFB fields. This is especially promising as, when we consider one PES, the singular values of the fields decay rapidly, albeit slower than in Fig.~\ref{fig:skyrme-fields-svd-combined}. The application of this method to fission PESs is the subject of a future work.

\begin{figure}[h]
    \centering
    \includegraphics[width=\linewidth]{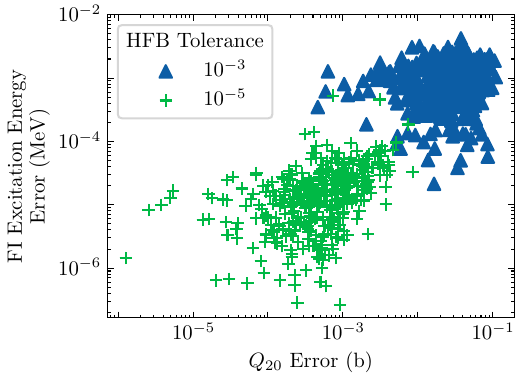}
    \caption{The error in the FI quadrupole moment $Q_{20}$ and excitation energy for $^{236}$Pu with 25 oscillator shells, at the LN level.}
    \label{fig:fission-isomer-error}
\end{figure}

\subsubsection{The Fission Isomer}
As a final demonstration, we consider the fission isomer (FI) of $^{236}$Pu. The FI is a highly elongated, short-lived state commonly found in actinide and superheavy nuclei~\cite{Metag1980,Thirolf2002,Walker2020,Singh2002}, typically with an excitation energy of only a few MeV. In the Wentzel–Kramers–Brillouin description of fission in actinide nuclei, fission tunneling pathways preferentially pass through the FI~\cite{Schunck2016,Flynn2022,Lay2024a}. In this sense the FI is a physically observable state that can be used to adjust fission properties of a particular model, in contrast to the fission barrier that was used in the SkM$^*$~\cite{Bartel1982} and D1S~\cite{Berger1984,Berger1991} EDFs.

Indeed, FIs for four actinide nuclei were included in the calibration of the UNEDF1 functional, however $^{236}$Pu was not. Experimentally, two FI for $^{236}$Pu are reported, at between 3 and 4 MeV~\cite{Singh2002}. These states have been little studied - for instance, the quadrupole moment is not reported for the 3 MeV isomer, and neither is an uncertainty estimate on the excitation energy. As such, a direct comparison with experimental data is out of the scope of this work.

We construct one emulator for the ground state, and a second for the FI. Both cases use unconstrained HFB calculations, so that the deformation of the isomeric state varies between EDF parameters. Again, we take as training 100 randomly sampled EDF parameters, using the same parameters for both the g.s. and FI emulators, and use the remaining as validation. Fig.~\ref{fig:fission-isomer-error} demonstrates the error on both the FI excitation energy and the quadrupole moment $Q_{20}$. We vary the solver tolerance on the original HFB solutions, considering both $10^{-3}$ and $10^{-5}$. As was discussed with Fig.~\ref{fig:error-vs-basis-size}, the emulator is only precise to within the tolerance of the training data. Hence, for the looser solver tolerance, the error is larger for both quantities. 

Even for the looser tolerance, the excitation energy is reproduced to within approximately 10 keV, which is not surprising from the results shown in Fig.~\ref{fig:error-vs-basis-size}. The quadrupole moment is, however, reproduced worse than the binding energy, with a relative error of $\sim10^{-3}$ for $Q_{20}$, as opposed to $\sim10^{-7}$ for the energy (the true values for each are approximately 80 b and 1.7 GeV, respectively). The relative error for higher even-parity moments is similar, while the odd-parity moments vanish identically.

This error is as expected from the previous accuracy discussion, namely that the EIM is more precise on the HFB energy than on other observables. Nevertheless, the emulator quality may be sufficient for a particular quantity of interest even for a loose SVD tolerance - for instance, the hybrid approach to estimating primary fission fragment yields from Refs.~\cite{Sadhukhan2020,Sadhukhan2022,Lay2024a} is somewhat robust to the particulars of the nuclear density. On the other hand, calibrations involving the root-mean-square proton radius tend to reproduce experimental values to within 0.02 fm, corresponding to a relative error on the order of $10^{-2}-10^{-3}$~\cite{Kortelainen2010,Kortelainen2012}. In the latter case, the tighter-tolerance HFB solutions should be used, so that the emulator error is still less than the calibration uncertainty.

Thus, we conclude that the output of the EIM is useful for simultaneous evaluation of multiple quantities that depend on the same HFB wave function. This is beneficial for calibration efforts, in which a number of observables for the same nucleus are used, e.g. binding energies and proton radii~\cite{Kortelainen2010}. It is also useful for fission calculations, where the collective inertia tensor depends on derivatives of the one-body density matrix~\cite{Schunck2016}, and fragment properties use the nuclear densities as input both in the hybrid approach of Refs.~\cite{Sadhukhan2020,Sadhukhan2022} and in time-dependent studies~\cite{Schunck2022}. Thus, the EIM is broadly useful for propagating statistical uncertainties through HFB calculations.

\section{Conclusion}\label{Sec: conclusions}

In this work, we demonstrated an emulator for axial HFB calculations based on the empirical interpolation method (EIM)~\cite{Barrault2004,Chaturantabut2010,Maday2016,Quarteroni2015,Benner2017,Brunton2019}. We have shown that the EIM emulator requires little training data to achieve keV-precision on the binding energy prediction. Further, this error level is tunable, and can be made arbitrarily precise. This is achieved with an order-of-magnitude speedup over the original HFB solver. We have also shown that this emulator works across the nuclear chart and is able to extrapolate in parameter space. Finally, we have shown that other quantities dependent on the one-body densities are also reproduced to high precision. Given that the core of the method is essentially linearization of non-affine operators, we anticipate this approach being broadly useful in existing self-consistent solvers~\cite{Perez2017,Niksic2014,Schunck2012}, although the resulting speedup should be evaluated on a case-by-case basis.

While EIM provides a considerable, useful, speedup over HFB calculations, it does not yet enable EDF calibration on a personal computer~\cite{Giuliani2023}. Thus, further improvements are desired. Some possible directions include: a hybrid emulation scheme, with the EIM of this work generating the training data for (e.g.) a neural-network emulator; an operator learning approach using genetic programming~\cite{Bakurov2025} or parametric matrix models~\cite{Cook2025,jin2025surrogate} to learn dynamical equations for the expansion coefficients as the EDF parameters vary; and using a projection-based scheme to determine an approximate orbital-free EDF, similar to the deep learning approach from Ref.~\cite{Hizawa2023}.

Beyond further emulator developments, there are numerous possible applications in existing fission studies. Certainly, the EIM is useful for uncertainty propagation in constrained calculations, which will enable routine uncertainty propagation in fission studies, beyond (and in addition to) comparing between a sample of EDFs. In particular, this will be useful for further probing multimodality in fission~\cite{Lay2024a}.

Preliminary work suggests that EIM, in fact, remains a useful tool even when changing the nuclear shape in constrained calculations. Thus, we anticipate EIM being useful when computing high-dimensional PESs to capture the simultaneous impact of triaxiality and pairing fluctuations~\cite{Matheson2019,Flynn2022,Sadhukhan2014}, interpolating PESs between nuclei for rapid neutron capture process studies~\cite{Lay2024b,Giuliani2020}, and describing induced fission via finite temperature DFT~\cite{Schunck2015prc}. Thus, this emulation approach is an important step in large-scale calculations of fission properties with quantified uncertainties.

\section*{Acknowledgments}
We wish to thank Edgard Bonilla and Kyle Beyer for useful discussions.

This work was supported in part by the U.S. Department of Energy under Award Numbers DOE-DE-NA0004074 (NNSA, the Stewardship Science Academic Alliances program), the U.S. Department of Energy, Office of Science, Office of Nuclear Physics, under Contract No. DE-AC02-06CH11357, the NUCLEI SciDAC-5 program, and the Department of Energy (DOE) Office of Science grant DE-SC0026198 (STREAMLINE 2 Collaboration).

\section*{Data Availability}
The following data is available at~\cite{Lay2026data}: the original HFB runs, and the text output from HFBTHO; the text output from the reduced GP and HFB solvers; and metadata for the various emulation schemes (e.g. run time for each sample point). Intermediate steps (e.g. wave function SVDs) tend to produce large files and so are not included; they will be made available on request.

\bibliographystyle{apsrev4-2}
\bibliography{references}

\typeout{get arXiv to do 4 passes: Label(s) may have changed. Rerun}

\end{document}